\input harvmac
\overfullrule=0pt
\def\Title#1#2{\rightline{#1}\ifx\answ\bigans\nopagenumbers\pageno0\vskip1in
\else\pageno1\vskip.8in\fi \centerline{\titlefont #2}\vskip .5in}

\font\ticp=cmcsc10
%
%
\def\sq2{\sqrt{2}}

\def\apm{\alpha'}

\def\s42{ 2^{-{1\over 4} } }

\def\g{\gamma}

\def\a{\alpha}

\def\cg2{\cos (\pi V)}
\def\sg2{\sin (\pi V)}
\def\cb2{\cos (\delta/2)}
\def\sb2{\sin (\delta/2)}

\font\cmss=cmss10 \font\cmsss=cmss10 at 7pt
\def\IZ{\relax\ifmmode\mathchoice
   {\hbox{\cmss Z\kern-.4em Z}}{\hbox{\cmss Z\kern-.4em Z}}
   {\lower.9pt\hbox{\cmsss Z\kern-.4em Z}}
   {\lower1.2pt\hbox{\cmsss Z\kern-.4em Z}}\else{\cmss Z\kern-.4emZ}\fi}

\def\apm{\alpha^{\prime}}

\def\a{\alpha}

\def\sa{r_0^2 {\rm sinh}^2\alpha }
\def\sg{r_0^2 {\rm sinh}^2\gamma }
\def\ss{r_0^2 {\rm sinh}^2\sigma }

\def\cg{r_0^2 {\rm cosh}^2\gamma }

\def\[{\left [}
\def\]{\right ]}
\def\({\left (}
\def\){\right )}
%
%
\lref\grla{ R. Gregory and R. Laflamme, Phys. Rev. Lett. {\bf 70} (1993)
2837; Nucl. Phys. {\bf B428} (1994) 399.}
\lref\jpup{J. Polchinski, private communication.}
\lref\rk{S. Ferrara and R. Kallosh, hep-th/9602136.}
\lref\host{G. Horowitz and A. Strominger, Nucl. Phys. {\bf B360} (1991) 197.  }
\lref\hhs{J. Horne, G. Horowitz, and A. Steif, Phys. Rev. Lett. {\bf 68} (1992)
 568, hep-th/9110065.  }
\lref\gkp{S. Gubser, I. Klebanov and A. Peet, hep-th/9602135.}
\lref\asup{A. Strominger, unpublished.}
\lref\witvar{E. Witten, Nucl. Phys. {\bf B 443} (1995) 85, hep-th/9503124.  }
\lref\cd{M. Cvetic and D. Youm, hep-th/9603100.}
\lref\dbr{J. Polchinski, S. Chaudhuri, and C. Johnson, hep-th/9602052.}
\lref\jp{J. Polchinski, hep-th/9510017.}
\lref\dm{S. Das and S. Mathur, hep-th/9601152.}
\lref\witb{E. Witten, hep-th/9510135.}
\lref\vgas{C. Vafa, hep-th/9511088.}
\lref\ghas{G. Horowitz and A. Strominger, hep-th/9602051.}
\lref\bsv{M. Bershadsky, V. Sadov and C. Vafa,
hep-th/9511222.}
\lref\vins{C. Vafa, hep-th/9512078.}
\lref\cvetd{M. Cvetic and D. Youm, hep-th/9507090.}
\lref\chrs{D. Christodolou, Phys. Rev. Lett. {\bf 25}, (1970) 1596;
D. Christodolou and R. Ruffini, Phys. Rev. {\bf D4}, (1971) 3552.}
\lref\cart{B. Carter, Nature {\bf 238} (1972) 71.}
\lref\penr{R. Penrose and R. Floyd, Nature {\bf 229} (1971) 77.}
\lref\hawka{S. Hawking, Phys. Rev. Lett. {\bf 26}, (1971) 1344.}
\lref\sussc{L.~Susskind,  Phys. Rev. Lett. {\bf 71}, (1993) 2367;
L.~Susskind and L.~Thorlacius, Phys. Rev. {\bf D49} (1994) 966;
L.~Susskind, ibid.  6606.}
\lref\polc{J. Dai, R. Leigh and J. Polchinski, Mod. Phys.
Lett. {\bf A4} (1989) 2073.}
\lref\ascv{A. Strominger and C. Vafa, hep-th/9601029.}
\lref\hrva{P. Horava, Phys. Lett. {\bf B231} (1989) 251.}
\lref\cakl{C. Callan and I. Klebanov, hep-th/9511173.}
\lref\prskll{J. Preskill, P. Schwarz, A. Shapere, S. Trivedi and
F. Wilczek, Mod. Phys. Lett. {\bf A6} (1991) 2353. }
\lref\sbg{S. Giddings, Phys. Rev {\bf D49} (1994) 4078.}
\lref\cghs{C. Callan, S. Giddings, J. Harvey, and A. Strominger,
Phys. Rev. {\bf D45} (1992) R1005.}
\lref\cvyo{M. Cvetic and D. Youm, hep-th/9507090.}
\lref\tse{A. Tseytlin, hep-th/9601117.}
\lref\cvpt{M. Cvetic, private communication.}
\lref\bhole{G. Horowitz and A. Strominger,
Nucl. Phys. {\bf B360} (1991) 197.}
\lref\bekb{J. Bekenstein, Phys. Rev {\bf D12} (1975) 3077.}
\lref\hawkb{S. Hawking, Phys. Rev {\bf D13} (1976) 191.}
\lref\wilc{P. Kraus and F. Wilczek, hep-th/9411219, Nucl. Phys.
{\bf B433} (1995) 403. }
\lref\ght{G. Gibbons, G. Horowitz, and P. Townsend, hep-th/9410073,
Class. Quantum Grav.,
{\bf 12} (1995) 297. }
\lref\intrp{G. Gibbons and P. Townsend, Phys. Rev. Lett.
{\bf 71} (1993) 3754.}
\lref\gmrn{G. Gibbons, Nucl. Phys. {\bf B207} (1982) 337;
G. Gibbons and K. Maeda Nucl. Phys. {\bf B298} (1988) 741.}
\lref\bch{J. Bardeen, B. Carter and S. Hawking,
Comm. Math. Phys. {\bf 31} (1973) 161.}
\lref\stas{A.~Strominger and S.~Trivedi,  Phys.~Rev. {\bf D48}
 (1993) 5778.}
\lref\jpas{J.~Polchinski and A.~Strominger,
hep-th/9407008, Phys. Rev. {\bf D50} (1994) 7403.}
\lref\send{A. Sen, hep-th/9510229, hep-th/9511026.}
\lref\cvet{M. Cvetic and A. Tseytlin, hep-th/9512031.}
\lref\kall{R. Kallosh, A. Linde, T. Ortin, A. Peet and 
A. van Proeyen, Phys. Rev. 
{\bf D46} (1992) 5278.}
\lref\lawi{F. Larsen and F. Wilczek, hep-th/9511064.}
\lref\bek{J. Bekenstein, Lett. Nuov. Cimento {\bf 4} (1972) 737,
Phys. Rev. {\bf D7} (1973) 2333, Phys. Rev. {\bf D9} (1974) 3292.}
\lref\hawk{S. Hawking, Nature {\bf 248} (1974) 30, Comm. Math. Phys.
{\bf 43} (1975) 199.}
\lref\cama{C. Callan and J. Maldacena, hep-th/9602043.}
\lref\sen{A. Sen, hep-th/9504147, Mod. Phys. Lett. {\bf A10} (1995) 2081.}
\lref\suss{L. Susskind, hep-th/9309145.}
\lref\sug{L. Susskind and J. Uglum, hep-th/9401070, Phys. Rev. {\bf D50}
 (1994) 2700.}
\lref\peet{A. Peet, hep-th/9506200.}
\lref\tei{C. Teitelboim, hep-th/9510180.}
\lref\carl{S. Carlip, gr-qc/9509024. }
\lref\thoo{G. 'tHooft, Nucl. Phys. {\bf B335} (1990) 138
Phys. Scr. {\bf T36} (1991) 247.}
\lref\fks{S. Ferrara, R. Kallosh and A. Strominger, hep-th/9508072,
Phys. Rev. {\bf D 52}, (1995) 5412 .}
\lref\spn{J. Breckenridge, R. Myers, A. Peet and C. Vafa, hep-th/9602065.}
\lref\vbd{J. Breckenridge, D. Lowe, R. Myers, A. Peet, A. Strominger 
and C. Vafa, hep-th/9603078.}
\lref\townsend{ C. Hull and P. Townsend, Nucl. Phys. {\bf B 438} (1995) 109,
hep-th/9410167. }
\lref\scherk{ E. Cremmer, J. Scherk and J. Schwarz, Phys. Lett. {\bf B 84 }
(1979) 83.}
\lref\dgl{ M. Douglas, hep-th/9512077.}
\lref\jmas{ J. Maldacena and A. Strominger, hep-th/9603060.}  
\lref\cjrm{C. Johnson, R. Khuri and R. Myers, hep-th/9603061.}
\lref\hlm{ G. Horowitz, D. Lowe and J. Maldacena, in preparation. }
\lref\cvetic{ M. Cvetic and D. Youm, hep-th/9508058.}

%
\Title{\vbox{\baselineskip12pt
\hbox{hep-th/9603109}}}
{\vbox{
\centerline {Nonextremal Black Hole Microstates}
\centerline{and U-Duality}}}
\centerline{{\ticp Gary T. Horowitz$^\dagger$, 
Juan M. Maldacena$^{\natural}$ and Andrew Strominger$^\dagger$}}
\vskip.1in
\centerline{$^\dagger$\it Department of Physics, University of California,
Santa Barbara, CA 93106, USA}
\centerline{\it gary@cosmic.physics.ucsb.edu \ \ \ andy@denali.physics.ucsb.edu}
\vskip.1in
\centerline{$^{\natural}$\it Joseph Henry Laboratories, Princeton University,
Princeton, NJ 08544, USA}
\centerline{\it malda@puhep1.Princeton.edu}
\bigskip
\centerline{\bf Abstract}
A six-parameter family of five-dimensional black hole solutions 
is constructed which are labeled by their mass, 
two asymptotic scalar fields and three charges. 
It is shown that the 
Bekenstein-Hawking entropy is exactly matched, arbitrarily far from 
extremality,  by a simple but mysterious 
duality-invariant extension of previously 
derived formulae for the number of D-brane states in string theory. 
\Date{}
%

\newsec{Introduction}

Recently a precise accounting of the microstates responsible 
for the Bekenstein-Hawking entropy 
of certain extremal BPS black holes has been given in string
theory \refs{\ascv \spn \jmas -\cjrm}.
Since black holes are nonperturbative objects, the calculations
required considerable understanding  of 
non-perturbative string theory and a certain class of solitons known as
D-branes  \jp.
Nevertheless, due to the special 
character of the BPS states involved, there is a sense in which
these calculations ``had to work''. 
The (weighted) 
number of 
BPS states is a topological invariant and 
so a string calculation at weak coupling can be compared to a 
semiclassical calculation at strong coupling. It would have been 
very strange indeed had the two calculations yielded 
different results. 
A failure of the 
string counting of states to match the Bekenstein-Hawking entropy 
would have been a serious blow to the notion that string theory 
is a complete quantum theory of gravity. 

Following \ascv\ several attempts were made \refs{\cama \ghas\ - \vbd} to 
extend these results to leading order 
away from extremality. These endeavors were on much 
shakier ground, because the number of non-BPS states 
is not topologically protected, and strong coupling effects 
could ruin the ability to extrapolate from the weakly-coupled 
stringy regime 
in which calculations are possible to the semiclassical regime in which
the black hole picture is valid. Nevertheless it was argued that 
strong coupling effects could be avoided in certain corners of the parameter 
space sufficiently near to extremality. The striking 
agreement discovered in \refs{\cama \ghas\ - \vbd} between the string and 
black hole calculations indicate that under favorable circumstances 
strong coupling effects can indeed be avoided\foot{However, as a note of 
caution, a similar analysis of near-extremal threebrane states in ten 
dimensions fails to match by a very puzzling $4/3$\refs{\gkp,\asup}.}.

In this paper we construct a six-parameter family of
five-dimensional black hole solutions with
arbitrary mass, three charges and arbitrary asymptotic values of 
two 
scalar fields.
The familiar Reissner-Nordstrom solution is included as
a special case. These black holes may be uniquely decomposed into 
a collection of D-branes, anti-D-branes and strings, whose numbers 
we denote 
($N_1,~N_{\bar 1},~N_5,~N_{\bar 5},~n_R,~n_L$). (An anti-D-brane is simply
a D-brane which is oriented in the opposite direction, and hence carries
the opposite sign of the RR charge.) These 
numbers are defined in section 2.4 
by matching thermodynamic properties of the black hole 
(under variation of the asymptotic parameters) to the  thermodynamic 
properties of a collection of ($N_1,~N_{\bar 1},~N_5,~N_{\bar 5},~n_R,~n_L$)
non-interacting branes, anti-branes and strings. In terms of these numbers 
the entropy takes the surprisingly simple, but mysterious, form 
\eqn\smira{S= 2 \pi( \sqrt{ N_1} + \sqrt{  N_{ \bar 1}}) 
( \sqrt{ N_5} + \sqrt{  N_{ \bar 5}})( \sqrt{ n_L} + \sqrt{ n_R})~.} 
Expressed in these variables the entropy is 
independent of both the string coupling and the internal five-volume.
This expression reduces, in several different limits, 
to expressions derived in several different weak coupling expansions 
in \cama, \ghas. It is also the simplest duality-invariant extension of 
those expressions.

The most surprising feature of \smira\ is that it is an exact 
expression which is valid arbitrarily far from extremality, even for 
large neutral Schwarzschild black holes. 
We will derive \smira\ 
from the stringy D-brane picture in several different weakly coupled 
limits, and motivate 
the full expression from duality.
We have not been able to obtain a 
stringy derivation of the full expression. This would require more than 
a weak-coupling analysis and seems to be 
quite challenging. The fact that the entropy 
is a product, rather than a sum, of terms suggests that 
the structure of the black hole Hilbert space may be quite different 
than anything previously imagined. At the very least it indicates that 
black holes and string theory have more lessons in store for us. 

In section 2 we describe the black hole solutions and discuss their properties.
The $N$'s are introduced and the above formula for the entropy is derived.
The D-brane analysis is given is section 3 and some possible extensions
of our results are discussed 
in section 4.

\newsec{The Black Hole Picture }
\subsec{The Solutions}
The low-energy action for ten-dimensional type IIB string theory
contains the terms
\eqn\fds{{1\over 16 \pi G_{10}}
\int d^{10}x \sqrt{- g} \[ R-{1 \over 2}(\nabla \phi )^2
-{1 \over 12} e^{\phi}H^2 \]}
in the ten-dimensional Einstein frame. $H$ denotes the RR three form field
strength, and $\phi$ is the dilaton. The NS three form, self-dual five
form, and second scalar are set to zero.
We will let $g$ denote
ten-dimensional string coupling and define the zero mode of  
$\phi$ so that 
$\phi$ vanishes asymptotically. The ten-dimensional Newton's 
constant is then $G_{10}=8 \pi^6 g^2$ with $\apm=1$. 
The metric in \fds\ differs from the string metric by a factor 
of $e^{\phi/2}$.
We wish to consider toroidal compactification to 
five dimensions with an $S^1$ of length $2\pi R$,  
a $T^4$ of four-volume $(2\pi)^4 V$, and momentum along the $S^1$.\foot{With
these conventions, T-duality sends $R$ to $1/R$ or $V$ to $1/V$, and S-duality
sends $g$ to $1/g$.}
This implies that the  metric takes the form
\eqn\lml{ds_{10}^2= e^{2\chi}\ dx_i dx^i + e^{2\psi}
(dx_5+A_\mu dx^\mu)^2 + e^{-2(4\chi + \psi)/3} ds_5^2}
where $\mu=0,1,...4$,\ $i = 6,...,9$, and all fields depend only on $x^\mu$.
The factor in front of $ds_5^2$ insures that this is the Einstein metric in
five dimensions.
We will assume that $x_5$ is periodically identified with period $2\pi R$,
$x_i$ are each identified with period $2\pi V^{1/4}$, and $\chi, \psi$  go
to zero asymptotically. 

A five parameter solution of the equations of motion following from
\fds\  was considered in \ghas\ which
was labeled by the energy, three charges, and $R$. The volume $V$ was
fixed in terms of the charges by the requirement that the field
$\chi$ in \lml\ remains constant. We now
remove that restriction and present a six parameter family of 
solutions\foot{After completion of this work, reference \cd\ 
appeared in which these solutions are also constructed.}. 
These can be obtained by first applying a $U$-duality transformation 
(discussed in section 2.3) to the
solution in \ghas\ which permutes the three charges \cama, and then 
applying a boost. 
The resulting  ten dimensional solution is given by
\eqn\dil{ e^{-2\phi } = \(1+   { \sg \over r^2 }\)\(1 + {\sa\over r^2 }\)^{-1}}
\eqn\metric{\eqalign{
ds^2 = &
 \( 1 + { \sa \over r^2}\)^{-3/4} \( 1 + { \sg \over r^2}\)^{-1/4} 
\left[ - dt^2 +dx_5^2
\right. \cr
+& \left. {
r^2_0  \over r^2} (\cosh \sigma dt + \sinh\sigma dx_5)^2
 +\( 1 + {\sa \over r^2}\) dx_i dx^i \] \cr
 +& \( 1 + { \sa \over r^2}\)^{1/4}\( 1 + { \sg \over r^2}\)^{3/4} \left[
\(1-{r_0^2 \over r^2}\)^{-1} dr^2 + r^2 d \Omega_3^2 \right]
}}
This solution is parameterized by the six independent quantities 
$\alpha,~~\g,~~\sigma,~~r_0,~~R$ and $V$. These may be traded 
for three charges, the mass, $R$ and $V$. The three charges are
\eqn\charges{
\eqalign{
   Q_1 &= {V\over 4\pi^2 g}\int e^\phi *H 
   = { V r_0^2  \over 2 g } \sinh 2 \a , \cr
   Q_5 &= {1\over 4\pi^2 g} \int H  =  { r_0^2\over 2g} \sinh 2 \g ,
\cr
  n &= {  R^2V r_0^2 \over 2  g^2} \sinh 2 \sigma ,
}}
where $*$ is the Hodge dual in the six dimensions $x^0,..,x^5$. 
The last charge $n$ is related to the  momentum around the $S^1$
by $P=  n/R$. All charges are normalized to be integers.
The energy is 
\eqn\mss{E=  {   R V r_0^2  \over 2  g^2} 
(\cosh 2 \alpha + \cosh 2 \gamma + \cosh 2 \sigma  )
}

If one reduces to five dimensions using \lml, the solution takes
the remarkably simple and symmetric form:
\eqn\solnfd{ds_5^2 =  - f^{-2/3} \(1-{r_0^2 \over r^2}\) dt^2 + f^{1/3} 
\[\(1-{r_0^2 \over r^2}\)^{-1} dr^2 + r^2 d \Omega_3^2 \right]}
where
\eqn\deff{ f= \(1+{\sa\over r^2} \)\(1+{\sg\over r^2} \)\(1+{\ss\over r^2} \)}
This is just the five-dimensional Schwarzschild metric with the time
and space components rescaled by different powers of $f$. The factored
form of $f$ was known  to hold for extremal solutions\refs{\tse,\cvpt}. It is surprising that
it continues to hold even in the nonextremal case. The solution is
manifestly invariant under permutations of the three boost parameters as
required by U-duality (see section 2.3).
The event horizon is clearly at $r=r_0$.
The coordinates we have used present the solution
in a simple and symmetric form, but they do not always cover the entire
spacetime. When all three charges are nonzero, the surface $r=0$ is
a smooth inner horizon. This is analogous to the situation in four
dimensions with four charges \cvetic . When at least one of the charges
is zero, the surface $r=0$ becomes singular.

Several thermodynamic quantities can be associated to 
this solution. They can be computed in either the ten dimensional or
five dimensional metrics and yield the same answer. For example, the
ADM energy \mss\ is the total energy of either solution.
The Bekenstein-Hawking entropy is 
\eqn\entropy{
S = {A_{10}\over 4 G_{10}} = {A_5\over 4 G_5} =
{ 2 \pi  R V  r_0^3 \over  g^2 }  \cosh \alpha \cosh \gamma \cosh \sigma.
}
where $A$ is the area of the horizon, and
we have used the fact that $ 8 \pi^6 g^2 = G_{10} = G_5 (2\pi)^5 R V $.
The Hawking temperature is 
\eqn\thwk{T= {1 \over 2 \pi  r_0 \cosh\alpha \cosh \gamma \cosh \sigma } .}
In ten dimensions, the black hole is characterized by  
pressures which describe how the energy changes for isentropic
variations in $R$ and $V$. In five dimensions, these are `charges' associated
with the two scalar fields. In either case, they are directly related to
the asymptotic fall-off of $\psi$ and $\chi$ in \lml\
and are given by
\eqn\psses{
\eqalign{
P_1=& 
{ R V r_0^2 \over 2  g^2} \left[ \cosh 2 \sigma -  {1\over 2} ( 
\cosh 2 \a + \cosh 2 \g ) \right]
\cr
P_2 =&
{ R V r_0^2\over 2  g^2} ( \cosh  2 \a -\cosh 2\g )
}}

The extremal limit corresponds to the limit $ r_0 \rightarrow 0$ 
with at least one of the boost parameters 
$\alpha , \gamma , \sigma \rightarrow \pm \infty $ keeping
$R, \ V$ and the associated charges \charges\ fixed.
If we keep all three charges nonzero in this limit, one obtains 
\eqn\extl{\eqalign{E_{ext} &= {R |Q_1|\over  g}  + 
{  R V |Q_5|\over  g}  +
 { |n|  \over R}  ~,\cr
S_{ext}&=2\pi\sqrt{|Q_1Q_5n|}~,\cr
            T_{ext}&=0~,\cr 
 P_{1ext}&= { |n| \over R }  - { R |Q_1|\over 2  g } -
 {R V |Q_5| \over 2 g }~, \cr
P_{2ext}& = {R |Q_1 | \over  g}  - { R V |Q_5|\over  g }  ~.}}
The first equation is the Bogomolnyi bound for this theory.

\subsec{Special cases}

The solution \dil,  \metric\ contains many well known solutions as special
cases. 
For example, suppose all three boost parameters are equal: $\a = \g = \sigma$.
Then the dilaton
is constant, and the internal five torus is constant. Letting $\hat r^2 = r^2
+ \sa$, the metric \solnfd\ is immediately recognized as the five dimensional
Reissner-Nordstr\"om solution.
The five dimensional Schwarzschild metric corresponds to 
$\a = \g = \sigma = 0$.

Next, suppose $\g = \sigma = 0$, so the only nonzero charge is $Q_1$. 
Then the ten-dimensional metric becomes
\eqn\metricbs{\eqalign{
ds^2 = &
 \( 1 + { \sa \over r^2}\)^{-3/4}  
\[ -\(1 -  {r_0^2 \over  r^2 }\) dt^2 + dx_5^2 \] \cr
 +& \( 1 + { \sa \over r^2}\)^{1/4} \left[
\(1-{r_0^2 \over r^2}\)^{-1} dr^2 + r^2 d \Omega_3^2 + dx^i dx_i\]
}}
This is the Einstein metric for the six dimensional black string solution  
(cross $T^4$) \hhs.\foot{The metric in \hhs\ is $e^{-\phi/2} $ times the above 
Einstein metric.
This is the string metric of the associated S-dual solution in which
$H$ represents the NS three form.}

Now suppose only $\g$ is nonzero, so the solution has only $Q_5$ charge.
Then we obtain
\eqn\metricfb{\eqalign{
ds^2 = &
 \( 1 + { \sg \over r^2}\)^{-1/4} 
 \[- \(1 -  {r_0^2 \over  r^2 }\) dt^2 + dx_5^2+ dx_i dx^i\] 
\cr 
 &+\( 1 + { \sg \over r^2}\)^{3/4} \left[
  \( 1 - { r_0^2  \over r^2 }\)^{-1} dr^2
  + r^2 d \Omega_3^2 \right] 
}}
This is the black five-brane
solution of \host.\foot{Once again, the metric of \host\  is $e^{-\phi/2} $
times the above metric, and expressed in terms of the radial coordinate
$\hat r^2 = r^2 + r_0^2 \sinh^2\g$.}

If $\sigma$ is the only boost parameter which is nonzero, then the metric
becomes
\eqn\metricm{\eqalign{
ds^2 = &
\left[ - dt^2 +dx_5^2
+{r^2_0  \over r^2} (\cosh \sigma dt + \sinh\sigma dx_5)^2
 + dx_i dx^i \] \cr
+& \(1-{r_0^2 \over r^2}\)^{-1} dr^2 + r^2 d \Omega_3^2 
}}
This is just the five dimensional Schwarzschild metric cross $T^5$,  with
a boost in
the $x_5$ direction. Similarly, adding  nonzero $\sigma$ to either \metricbs\
or \metricfb\ corresponds to  adding 
a boost in the $x_5$ direction. Finally,
if $\a=\g$, then the dilaton and volume of $T^4$ remain constant. This is
just the dyonic black string solution discussed in \ghas\ with $r_+^2 = r_0^2
\cosh^2\g, \ r_-^2 = r_0^2 \sinh^2 \g$.

\subsec{Duality Transformations}

The full type IIB string theory compactified on $T^5$ yields $N=8$ supergravity
in five dimensions. This theory has a global $E_{6(6)}$ symmetry and contains
27 gauge fields which transform in the 
27 of $E_{6(6)}$, and 42  scalar fields which parameterize the coset
 $E_{6(6)}/Sp(8)$ \scherk . In string theory this symmetry is believed to be
 broken down to an 
 $E_{6(6)}(Z)$ duality group.
Since we have only kept three of the gauge fields, we will be interested
in an $S_3$ subgroup which permutes the 
three charges $(Q_1,Q_5,n)$\cama. This is generated by $A$ and $B$ 
which act as 
\eqn\dgr{\eqalign{
   A=T_6 T_7 T_8 T_9:~~~~Q_1'&=Q_5,~~~~~ g'={g\over R_6R_7R_8R_9},\cr
                                         Q_5'&=Q_1,~~~~~ R_5'=R_5,\cr
                                           n'&=n,  ~~~~~~
 R_i'={1 \over R_i}, ~i = 6,7,8,9,\cr~&\cr
                   B =  T_9 T_8 T_7 T_6 S T_6 T_5:~~~~
Q_1'&=Q_5,~~~~~~~ g'={g\over R_5 R_7 R_8 R_9},\cr
Q_5'&=n,~~~~~~~~  R_5 '={\sqrt{ R_6 \over g R_5} },\cr
n'&=Q_1,~~~~~~R_6'=\sqrt{ g R_6 \over R_5 },\cr
                            & ~~~~~~~~~~~~~~~ 
R'_i = \sqrt{ g \over R_5 R_6 }
{1\over R_i },~
i= 7,8,9  }}
where $R_i$ is the radius of the internal direction $i$, similarly $T_i$ is
T duality along the  internal direction $i$.  
Note that after the last transformation the string is pointing along a different
direction, we could use a simple rotation in the internal space to take it
to the original configuration. 

\subsec{Relation to Fundamental Strings and D-branes}

We now show that there is a formal sense in 
which the entire family of solutions
discussed in section 2.2 can 
be viewed as  ``built up" of branes, anti-branes, and strings.
The extremal limits of \metricbs, \metricfb, and \metricm, are obtained
by letting $r_0$ go to zero, the boost parameter go to infinity such that the 
charge is fixed. These extremal metrics represent a D-onebrane wrapping
the $S^1$, a D-fivebrane wrapping the $T^5$,
or the momentum mode of a 
fundamental string around the $S^1$. 
 From \mss\ and \psses\ we see that a single
 onebrane {\it or} anti-onebrane has
mass and pressures 
\eqn\obc{M={R\over  g}, \qquad P_1 = -{R\over 2g} \qquad P_2 = {R\over g}}
Of course a onebrane has $Q_1 =1$, while an anti-onebrane has $Q_1 = -1$.
A single fivebrane  or 
anti-fivebrane has
\eqn\fbc{M={RV\over  g}, \qquad P_1 = -{RV\over 2g} \qquad P_2 = - {RV\over g}}
For left- or right-moving strings 
\eqn\sbc{ M={1\over  R}, \qquad P_1 = {1\over R} \qquad P_2 = 0}

Given \obc\ - \sbc,  and the relations \charges, \mss, and \psses,
it is possible to trade the six parameters of the general solution 
for the six quantities ($N_1,~N_{\bar 1},~N_5,~N_{\bar 5},~n_R,~n_L$)
which are the ``numbers'' of onebranes, anti-onebranes, fivebranes, 
anti-fivebranes, right-moving strings and left-moving strings 
respectively. This is accomplished by equating the total 
mass, pressures and charges of the black hole with 
those of a collection of  ($N_1,~N_{\bar 1},~N_5,~N_{\bar 5},~n_R,~n_L$)
{\it non-interacting} ``constituent'' branes, antibranes and strings. 
By non-interacting we mean that the masses and pressures are simply the sums 
of the masses and pressures of the constituents. 
The resulting expression for the $N$'s are 
\eqn\dbranes{
\eqalign{
   N_1=& { V  r_0^2  \over 4g} e^{ 2 \a},
\cr
    N_{\bar 1}= & 
{ V  r_0^2  \over 4 g} e^{- 2 \a}
\cr
 N_5 =& { r_0^2 \over 4g} e^{ 2 \g },
\cr
 N_{\bar 5} =& 
{ r_0^2 \over 4g} e^{- 2 \g },
\cr
 n_R = & 
{   r_0^2 R^2 V  \over 4 g^2} e^{ 2 \sigma },
\cr
 n_L = & { r_0^2 R^2 V  \over 4 g^2} e^{ -2 \sigma }.
}
}
\dbranes\ is the {\it definition} of the $N$'s, but we will refer to them 
as the numbers of branes, antibranes and strings because 
(as will be seen) they reduce 
to those numbers in certain limits where these concepts are well defined. 

In terms of the numbers \dbranes, the charges are simply
$Q_1 = N_1 - N_{\bar 1}, \ Q_5 = N_5 - N_{\bar 5}, \ n = n_R - n_L$,
the total energy is
\eqn\energybr{ E = {R\over  g} (N_1 + N_{\bar 1}) + { RV \over  g} 
     (N_5 + N_{\bar 5}) + {1 \over R} (n_R + n_L)   }
and the volume  and radius are 
\eqn\forvol{ V =  \( {N_1 N_{\bar 1} \over 
N_5 N_{\bar 5}} \)^{1/2}}
\eqn\forlen{ R =  \( { g^2 n_R n_L \over N_1 N_{\bar 1} } \)^{1/4}}
 From \extl\ we see that the extremal solutions correspond to including
 either branes or anti-branes, but not both. Notice that for the general 
 Reissner-Nordstrom solutions ($\a = \g = \sigma$) the contribution to
 the total energy from onebranes, fivebranes, and strings are all equal:
\eqn\rnequal{ {R\over  g} (N_1 + N_{\bar 1})={ RV \over  g}
(N_5 + N_{\bar 5})={1 \over R} (n_R + n_L)   .}
The actual number of branes of each type  depends on $R$ and $V$ and
can be very different.

Of course there seems to be no reason for neglecting interactions 
between collections of branes and strings 
composing a highly non-extremal black hole at strong or intermediate 
coupling. Hence 
the definitions \dbranes\ would seem to be inappropriate for 
describing a generic black hole. 
However, the utility of these definitions can be seen when we 
reexpress the black hole entropy \entropy\ in terms of the $N$'s. 
It takes the remarkably simple form
 \eqn\smira{S= 2 \pi( \sqrt{ N_1} + \sqrt{  N_{ \bar 1}}) 
( \sqrt{ N_5} + \sqrt{  N_{ \bar 5}})( \sqrt{ n_L} + \sqrt{ n_R})~.}
In the next section we shall see that this
entropy formula 
arises 
naturally in the D-brane picture.

\newsec{The D-Brane Picture}

In this section we will describe and compute the entropy of 
a collection of D-branes and strings and reproduce the formula \smira\
in various limits. Throughout most of
this section we will presume that the ten-dimensional string 
coupling is extremely small, so that D-brane perturbation theory 
is accurate. This implies that the string scale is large in Planck
units and the ``black hole'' is surrounded 
by a stringy halo which is large compared to its Schwarzschild radius.
Hence the best physical description is as a bound collection of 
D-branes and strings rather than as a semiclassical black hole solution.

To begin, consider a BPS-saturated state 
consisting of $N_1$ onebranes, $N_5$ fivebranes,
$n_R$ right moving strings, and no anti-branes or left-moving momenta.
We have seen that this corresponds to an extremal black hole.
The  fivebranes 
are wrapped around the $T^5$  with size $VR$ and 
the onebranes are wrapped around the $S^1$ of radius $R$. If $N_5=1$,
the $N_1$ onebranes are marginally bound to the 
fivebrane \refs{\send,\vgas} but are 
free to move within the transverse $T^4$. This motion is generated 
by the  $(1,1)$ Dirichlet strings both of whose ends are stuck to the 
onebranes,
but can carry momentum along the $S^1$. (The $(1,5)$ strings carry charge 
and so are confined.) The extremal black hole with 
nonzero $n_R$-charge corresponds to a BPS state with nonzero momentum 
along the $S^1$, {\it i.e.} with right-moving but no left-moving 
Dirichlet strings. The number of such states for fixed $n_R$ follows 
from the standard thermodynamic formula
for the entropy of
$N_B$ ($N_F$) species of right-moving bosons (fermions)
with total energy $E_R$ in a box of length $L$
\eqn\est{S=\sqrt{\pi (2N_B+N_F)E_RL\over 6}~.}
Using $N_F=N_B=4N_1N_5$, $ L= 2\pi R$,
$E_{R}= n_R/R$,  
\est\ becomes
\refs{\ascv}
\eqn\ssr{
S= 2\pi\sqrt{ N_1N_5n_R}
}
which agrees with the formula \entropy\ for $N_{\bar 1}=N_{\bar 5}=n_L=0$.

For $N_5 >1$ the picture is somewhat different. In that case the 
number of $(1,5)$ strings $(N_1N_5)$ is greater than the rank of the
gauge group.
Their potential accordingly has $D$-flat directions and they can condense 
(they are five-dimensional hypermultiplets) 
and break all the gauge symmetries. However in this phase there are still 
$N_1N_5$ Dirichlet strings so the formula \ssr\ remains valid, as promised 
by duality. The best way to describe this situation is to represent the 
onebranes as instantons in the fivebrane 
\refs{\vins,\dgl}, but in the following we concentrate on the simplest case 
$N_5=1$. 

Since the above counting does not depend on the sign of the charges,
it is clear that if we had started with $N_{\bar 1}$ anti-onebranes and
no onebranes, the counting would have been identical with the result
\eqn\ssar{S= 2\pi\sqrt{ N_{\bar 1} N_5 n_R},}
for $N_{1}=N_{\bar 5}=n_L=0$.
Similarly, we could have started with anti-fivebranes or left moving
strings and obtained analogous formula. This shows that the entropy
\entropy\ correctly reproduces the number of states when one integer
in each of the three factors is zero, which are the extremal black holes.

We now consider small deviations from extremality (we assume
$N_{\bar 1} = N_{\bar 5} =n_L=0$ initially)
in three  regimes
 where one of the excitations is much lighter
than the other two.  In this case, the nonextremal 
entropy will come from adding only left moving
momentum or only  anti-onebranes or only anti-fivebranes.

Let us review first the case considered in \ghas\ which corresponds
to the case in which the momentum modes are light.
It can be seen from \energybr\ that this case corresponds to 
taking $R$ very large, we always consider 
 low
energies $\delta E$ 
above extremality (but still big enough so that
we have a large number of right movers $n_R \gg 1$).
In this case the D-branes get very heavy and 
the light modes of the system are  
left and right moving  excitations of the $(1,1)$ Dirichlet strings 
which carry energy $E_{R,L}= n_{R,L}/R$. For nonzero 
$\delta E$ both left- and right-moving modes will be present.
The rate of interactions between them is proportional to the string 
coupling $g$ as well as the density of strings, which is inversely proportional 
to $R$. Hence for fixed coupling interactions can be ignored to leading 
order in a $1/R$ expansion. The entropy is then just the sum 
of left- and right-moving contributions \ghas 
\eqn\sstr{
S= 2 \pi \sqrt{N_1N_5} (\sqrt{ n_R}  + \sqrt{ n_L })~
}
which again agrees with \smira\  for $N_{\bar 1}=N_{\bar 5}=0$.

A second way to get light modes is to take $R$ to be very small 
with $RV$ fixed. In 
that case momentum modes of strings and wrapping modes of fivebranes 
are heavy but winding modes of 
onebranes are light. The best way to analyze this is to 
T-dualize along the $S^1$ to a IIA theory with 
large $\tilde R=1/R$.\foot{We 
wish to keep the IIA string coupling fixed as $R$ gets small.}
The onebranes become zerobranes, the fivebranes become fourbranes
and momentum becomes fundamental string winding. Let us first 
reproduce the extremal entropy \ssr\ in this picture. For $N_1=0$, 
$N_5=1$ and arbitrary $n_R$ one has $n_R$ fundamental strings which 
wind once around the $S^1$ and have both ends stuck on the fourbrane.
To reproduce \ssr\ we must count the number of ways of adding 
zerobranes with total charge 
$N_1$. The zerobranes are like beads (in a zero-momentum wave function) 
threaded on any one of the 
$n_R$ fundamental strings. Since $V$ is large the strings are far apart and 
the beads can be threaded by only one string at a time. 
There is one species of bead for every 
integer value of the charge. (This was first postulated in \witvar\ 
as required for compatibility of string theory and eleven-dimensional
supergravity, and has subsequently been confirmed in a variety of contexts.)
Furthermore each such bead is an $N=2$ hypermultiplet with 4 bosonic 
and 4 fermionic
states. Supersymmetry of the extremal configuration requires that we use only 
positively charged beads. 
Counting the number of such states with total charge $N_1$ 
is isomorphic to counting the number of 
states of $4n_R$ bosonic and $4n_R$ fermionic oscillators with total energy 
$N_1$.
This reproduces exactly the extremal entropy \ssr. 
In the case $N_5 >1 $ we can think that the four branes are separated and
there are $N_5 $ spaces between them where $n_R$  strings can start and end 
on different points on
the four branes, so there are $N_5 n_R$ distinct strings where 
we can put the zero branes. 
Now let us consider 
nonzero $\delta E$. This requires both charges of beads. However for large 
$\tilde R$ (small $R$) the beads are dilute. 
Over most of phase space the forces between the 
beads are small and so interactions  can be ignored. 
The entropy is additive and given by 
\eqn\ssor{S= 2 \pi \sqrt{ N_5n_R}
( \sqrt{N_1} + \sqrt{  N_{\bar 1} } )~,}
where $N_1$ ($  N_{\bar 1}$) is the number of zerobranes (anti-zerobranes).
This is again in agreement with \smira\  for $N_{\bar 5}=n_L=0$.

A third way to get light modes is to keep $R$ fixed and take 
$V$ to be very small.
 In this case the fivebranes become 
light and dominate the near-extremal entropy. This is related by 
T-duality to the preceding case. The near-extremal entropy is then 
 \eqn\ssfr{S= 2 \pi \sqrt{N_1  n_R}
( \sqrt{N_5} + \sqrt{  N_{\bar 5} } )~,}
where $N_5$ ($ N_{\bar 5}$) is the number of zerobranes (anti-zerobranes).

Using the U-duality 
transformations \dgr\ 
the role of fivebranes, one branes and strings 
in the preceding can be permuted. A check on this 
is given by the fact that  
that  \sstr, \ssor\ and \ssfr\ are permuted into one another
under interchanges of fivebranes, onebranes and strings.

Away from extremality and 
the special limits of moduli space discussed above one expects all
six quantities $N_1, \bar N_1, N_5, \bar N_5, n_R$ and $n_L$ to 
be nonzero. 
There is a natural and simple expression for the entropy 
which reduces to \sstr \ssor \ssfr\
and also is invariant under the permutations as required by duality. 
That expression is 
\eqn\smira{S= 2 \pi( \sqrt{ N_1} + \sqrt{  N_{ \bar 1}}) 
( \sqrt{ N_5} + \sqrt{  N_{ \bar 5}})( \sqrt{ n_L} + \sqrt{ n_R})~.}
We do not know of a systematic derivation of this formula using 
D-brane technology.
However miraculously it agrees 
with the Bekenstein-Hawking entropy calculated in the previous section
from the area of the event horizon. 

 From the D-brane point of view,
if \smira\ is the correct formula 
and the $N$'s can be interpreted as the number of branes and anti-branes,
then there seems to be a  discrepancy in the number of free parameters: in
the D-brane picture, one can 
specify the six numbers in \smira\ plus $V$ and $R$. 
However, it turns out that if one maximizes the entropy \smira\
keeping the charges \charges\  mass \energybr\ and $R,\ V$ fixed, then the 
proportion of branes and anti-branes that results is
 precisely the amount  present in
the black hole solution \dbranes . This supports the 
picture of the black hole as an ensemble of branes in
thermodynamic equilibrium.\foot{Even in the black hole picture, $R$ and $V$
are not completely arbitrary, since (away from extremality)
if they are too large, the solution
becomes classically unstable \grla.}

\newsec{Discussion}

In principle one might hope to go beyond  the special 
weakly-coupled limits in \sstr\ -   \ssfr\ and derive 
\smira\ by counting D-brane configurations. The problem is that one 
immediately runs into strong coupling.  There are
interactions between
D-branes which are independent of the string coupling, e.g., since
$G \sim g^2$ and $E \sim 1/g$, gravitational interactions are independent of
$g$, and so cannot be suppressed by making $g$ small. 
For example consider the regime in which  $g$ is small,
$R$ is large 
and $V$ is order one. Then the string modes are weakly coupled 
and \sstr\ can be derived. However, the
onebrane-anti-onebrane pairs  are on 
top of each other with coupling of order one and would seem to 
immediately annihilate, even if $g$ is small.\foot{It 
 is conceivable that this annihilation is suppressed 
by some new strong coupling effects.} Hence it does not appear to be 
sensible to discuss onebrane-anti-onebrane pairs in this regime.
Nevertheless the simplicity of \smira\ begs for an explanation.

If one simply expands \smira\
one obtains a representation of the entropy of a nonextreme black hole
in terms of the sum of eight extremal black holes. This 
suggests that the entropy of a generic black hole might be 
explained as the sums of the entropies of extremal constituents. 
However, the eight
extremal black holes  have a total mass which is four times
the mass of the initial black hole (since each integer $N$ appears four 
times in the sum). So this approach cannot explain the entropy formula.

While it is probably not possible to derive all of \smira\ from known 
techniques 
it may be both possible and instructive to go beyond what we have done.
For small $R$ and fixed $V$ the couplings between 
both onebrane-anti-onebrane pairs and fivebrane-anti-fivebrane pairs are weak 
in the sense that (in the T-dual picture)
they are dilute and take a long time to annihilate. 
If we add a small amount of energy to the extremal configuration in this small
$R$ regime, it should go into configurations which look approximately 
like collections of strings,
onebranes, anti-onebranes, fivebranes and anti-fivebranes (note from 
\dbranes\ that $n_L \rightarrow 0$ for small $R$.). By counting 
such configurations one should be able to determine the ratio of 
anti-onebranes to antifivebranes as a function of the energy, $R$, $V$ 
and the three charges to leading order in $R$.
After transforming to the $N$'s, this would give a check on \smira\ with 
only $n_L=0$.

A puzzling feature of \smira\ is that it only involves onebranes,
fivebranes, and strings. This is understandable for extremal solutions
with these charges, but when one moves away from extremality, one might
expect pairs of threebranes and anti-threebranes or fundamental string
winding modes to contribute to the entropy. To see the contributions
of these other objects, one should start with the full Type II string theory
compactified on $T^5$. The  low energy limit of this theory 
is $N=8$ supergravity
in five dimensions. This theory has 27 gauge fields, 42 scalars and
a global $E_6$ symmetry. Since only the scalar fields which couple to the
gauge fields are nontrivial in a black hole background, we expect the general
solution to be characterized by 27 scalars in addition to the 27 charges.
(Since an overall shift of all the scalars can be compensated by
rescaling the metric, one can interpret the 27 scalar parameters as 26 scalars
plus the ADM energy.) 
 Each charge corresponds to a type of
soliton or string. Thus we expect the solution to again be characterized
by the number of solitons and anti-solitons. For an extremal black hole,
the entropy can be written in the $E_6$ invariant form \refs{\jpup, \rk}
\eqn\eeninv{ S= 2\pi |T_{ABC} V^A V^B V^C|^{1/2}  }
where $V^A$ is the 27 dimensional charge vector and $T_{ABC}$ is a 
symmetric cubic
invariant in $E_6$. For the nonextremal black holes, the above argument
suggests that one
can introduce two vectors $V_i^A \ i=1,2$ which represent the number
of solitons and anti-solitons. Although we have not done the calculation,
the general black hole entropy might take the
$E_6$ invariant form
\eqn\geninv{ S = 2\pi \sum_{i,j,k} |T_{ABC} V_i^A V_j^B V_k^C|^{1/2} }
where the sum is over the two possible values of each of $i,\ j$ and $k$.
If this is the case, the entropy of nonextremal
black holes could be represented in terms of solitons and anti-solitons
in many different (equivalent) ways which
are related by $E_6$ transformations.

The entropy of four dimensional black holes can also be expressed  in a 
form similar to \smira\ 
 \hlm , using the representation given in \refs{\jmas,\cjrm}.

{\bf Acknowledgments}
We would like to thank C. Callan, R. Kallosh, D. Lowe, 
J. Polchinski, and L. Susskind for useful discussions. 
The research of G.H. is supported in part by NSF Grant PHY95-07065,
 that of A.S. is supported in part by DOE grant DOE-91ER40618 and that
of J.M. is supported in part by DOE grant DE-FG02-91ER40671.

\listrefs
\end